\documentstyle[12pt]{article}

\newcommand{\be}{\begin{equation}}
\newcommand{\ee}{\end{equation}}
\newcommand{\ba}{\begin{eqnarray}}
\newcommand{\ea}{\end{eqnarray}}
\def\npb#1#2#3{{\it Nucl.\ Phys.} {\bf B#1} (19#2) #3}
\def\mpla#1#2#3{{\it Mod.\ Phys.\ Lett.} {\bf A#1} (19#2) #3}
\def\cmp#1#2#3{{\it Commun.\ Math. Phys.} {\bf #1} (19#2) #3}
\def\endli{\hfill\break}

\def\p{\partial}
\def\der{\,{\rm d}}
\def\winfty{w_\infty}
\def\su#1{{\rm SU}(#1)}
\def\u#1{{\rm U}(#1)}

\def\ints{\int_\Sigma\der^2\sigma\;}

\def\rth{\sqrt{h}\;}

\def\christ#1#2#3{\Gamma^{#1}_{#2#3}\,}
\def\e#1{{\rm e}^{#1}}
\def\CA{{\cal A}}                   \def\CV{{\cal V}}
\def\CL{{\cal L}}                   \def\CZ{{\cal Z}}
                   
\def\CD{{\cal D}}                   \def\CM{{\cal M}}
\begin{document}
\hsize37truepc\vsize52truepc
\hoffset=-.5truein\voffset=-0.8truein
\setlength{\baselineskip}{17pt plus 1pt minus 1pt}
\setlength{\textheight}{21.5cm}
\hfill \hbox{EFI-93-66\hspace{.39in}}

\hfill \hbox{November 1993}

\vskip.5truein

\leftline{\large\rm Topological Strings and QCD in Two Dimensions\footnote{
talk given at the Carg\`ese conference on ``Recent Developments in String
Theory, Conformal Field Theory and Topological Field Theory'' (May 1993), to
appear in the Proceedings}}
\vskip.65truein

\hbox{\obeylines\baselineskip12pt\parskip0pt\parindent0pt\hskip1.1truein
\vbox{\sc Petr Ho\v rava
\vskip.1truein
\it Enrico Fermi Institute
\it University of Chicago
\it 5640 S. Ellis Ave.
\it Chicago IL 60637, USA
}}
\vskip .65truein
\noindent
{\bf Abstract.}
I present a new class of topological string theories, and discuss them in two
dimensions as candidates for the string description of large-$N$ QCD.  The
starting point is a new class of topological sigma models, whose path integral
is localized to the moduli space of harmonic maps from the worldsheet to the
target.  The Lagrangian is of fourth order in worldsheet derivatives.  After
gauging worldsheet diffeomorphisms in this ``harmonic topological sigma
model,'' we obtain a topological string theory dominated by minimal-area
maps.  The bosonic part of this ``topological rigid string'' Lagrangian
coincides with the Lagrangian proposed by Polyakov for the QCD string in
higher dimensions.

\vskip.65truein

\centerline{\bf 1. Introduction}
\vskip.1truein

Strings undoubtedly represent one of the most interesting and universal
patterns in Nature, emerging at vastly different scales and in different
physical phenomena.  The Regge phenomenology of hadrons and the structure of
dual models were first indications that the adequate degrees of freedom
for the theory of strong interactions at low energies are qualitatively those
of a string theory.  These ideas received a more specific support soon
after QCD had been identified as the proper theory for strong interactions.
Two basic techniques that reveal some form of string structure in QCD are the
expansion in the number of colors, and the strong coupling expansion.
Yet, even though the success of QCD in the ultraviolet has been enormous,
any progress in the identification of the ``stringy'' fixed point that governs
the theory in the infrared has been very limited (for recent reviews, see
[1]).

In the meantime, strings have been recognized as playing an important,
although still mostly enigmatic, r\^ole in quantization of gravity.  Ever
since, string theory of quantum gravity has been changing profoundly the basic
concepts of space and time in quantum theory, and promises to do so even more
after it is understood better as a second-quantized theory.  Among the
fascinating tentative results in this direction are the apparent lack of
degrees of freedom in string theory at short distances,  an extremely soft
behavior of strings in super-Planckian collisions, and the existence of a
minimal length/enlarged uncertainty principle, to mention just a few.  The
idea of a possible topological phase of quantum gravity (and string theory) at
short distances also arised in this context.  Without any doubt, many aspects
of string theory have been elucidated in the context of quantum gravity.

At a more mundane level, however, one may return back to old unsolved
problems, and start wondering whether the enormous progress accomplished in
string theory during last decade could possibly lead to some new insight,
and help in the quest for the string theory of QCD in the infrared.  This
program has been raised recently by Gross [2] and Gross and Taylor [3], in the
context of the exactly solvable, pure-glue QCD in two dimensions.
\index{QCD!two dimensional} The idea is that the theory, despite having no
local degrees of freedom, is rich enough in the content that it leads to a
non-trivial string theory.  It is exactly solvable on any compact spacetime
manifold, thus supplying us with lots of information about the possible
character of the string theory.  The authors of [3] have been able to
interpret the results of the $1/N$ expansion for 2D QCD in terms of simple
rules for counting maps between surfaces (very similar results were obtained
for Wilson loops on the plane some time ago by Kazakov and Kostov, see [4]).
The next step is to try and find the worldsheet Lagrangian for the string
theory, with the hope that this two dimensional QCD string
\index{QCD string!two dimensional} Lagrangian could grasp some important
aspects of the ``stringy'' fixed point that supposedly describes the real case
of our interest, namely four dimensional QCD in the infrared.  This represents
an extremely interesting program for string theorists.

A successful identification of the worldsheet theory for two dimensional QCD
could also have implications for string theory itself.  Assuming that the
worldsheet theory for the 2D QCD string is found, we can follow standard rules
and construct its corresponding string field theory.  This theory could be
sufficiently complicated, but in this case -- unlike in the ``fundamental''
string theory of quantum gravity -- we do know the correct spacetime
description of the theory, this being given by the QCD Lagrangian itself!
Thus, the 2D QCD string theory could even teach us something about the
conundrums of string field theory for quantum gravity.

In this talk I present a new class of topological string theories, and discuss
them in two dimensions as candidates for the worldsheet description of the QCD
string theory.  The presentation is based on unpublished results that will
appear elsewhere [5].

In \S{2} I review very briefly some aspects of quantum Yang-Mills theory in
two dimensions, \index{Yang-Mills theory!two dimensional} its relation to
topological Yang-Mills theory, \index{Yang-Mills theory!topological} and
the interpretation of the $1/N$ expansion of the $\su{N}$ theory in terms of
maps between surfaces as obtained by the authors of [2-4].

In \S{3}, a new class of topological sigma models is defined, as the
``worldsheet matter'' for the topological string theory that we are about to
construct.  The path integral of these topological sigma models is localized
at the moduli space of harmonic maps from the worldsheet to the target
(with respect to a fixed metric on both). To distinguish the theory from the
traditional topological sigma model dominated by holomorphic maps, I refer to
it as the ``harmonic topological sigma model'' \index{Topological sigma
models!harmonic} henceforth.  The bosonic part of the Lagrangian is of fourth
order in worldsheet derivatives, and is reminiscent of the rigid string
\index{Rigid string} Lagrangian proposed by Polyakov [6] some time ago as a
candidate for the QCD string in higher dimensions.

In \S{4} we gauge worldsheet diffeomorphisms in the harmonic topological
sigma model.  Instead of introducing a dynamical metric on the worldsheet, we
will replace the fixed metric by the induced metric.  This completes the
analogy with the rigid string, and I refer to the theory as the ``topological
rigid string theory.'' \index{Topological rigid string theory}
\index{Rigid string!topological} The instanton moduli space that dominates the
path integral is the space of maps harmonic in their own induced metric,
i.e.\ the moduli space of minimal-area maps.

In \S{5} I discuss possible variations on the theme of topological rigid
strings in dimensions higher than two, in particular in four dimensions.

\vskip.2truein
\centerline{\bf 2. Yang-Mills Theory and Large-$N$ QCD in Two Dimensions}
\nopagebreak\vskip.1truein\nopagebreak

The recent revival of interest in two dimensional Yang-Mills theory has led
to very interesting results.  The quantum theory is exactly solvable on any
Riemann surface, a fact that can be explained from different points of view.
The original one, due to Migdal and recently Rusakov [7], uses a RG-invariant
lattice formulation.  Another point of view, developed by Witten [8], explains
the exact solvability in the continuum theory in terms of localization of the
path integral in equivariant cohomology/topological Yang-Mills theory to a
moduli space of classical configurations.  Since we are interested in the
continuum limit of the worldsheet theory for the QCD string, the results of
Witten seem to be particularly relevant.  In [8] it is shown that after some
fermion fields are added to the Yang-Mills theory, the ``physical'' Yang-Mills
Lagrangian can be treated as a BRST invariant observable in the underlying
topological Yang-Mills theory.  This correspondence then allows one to
calculate the partition function of the ``physical'' theory as a correlation
function in the underlying topological theory.  In this picture, two
dimensional Yang-Mills theory is so special because there is an underlying
topological theory that actually governs the calculation of the partition
function in the ``physical'' theory.

The partition function of the Yang-Mills theory on an arbitrary two
dimensional surface $M$ of genus $G$ is given by
\be
\CZ\equiv\int\CD A_\mu\;\e{(1/\tilde g^2)\int_M\der\mu\,{\rm Tr}\,F_{\mu\nu}
F^{\mu\nu}}=\sum_R({\rm dim}\, R)^{2-2G}\e{-\tilde g^2AC_2(R)/2},
\label{eepartit}
\ee
where $\der\mu$ is the Riemannian volume element and $A$ the area of a fixed
metric $G_{\mu\nu}$ on $M$, the sum runs over all irreducible representations
of the gauge group, and $C_2(R)$ is the second Casimir of $R$.  For $\su{N}$
we can take the large-$N$ limit that keeps $\lambda\equiv\tilde g^2N$ fixed,
and expand the exact result (\ref{eepartit}) in the powers of $1/N$.  In
[2,3], Gross and Taylor obtained the following expression for the $1/N$
expansion of (\ref{eepartit}):
\be
\CZ=\sum_{g\geq 0}\frac{1}{N^{2g-2}}\sum_{n,\tilde n}\e{-(n+\tilde n)\lambda
A/2}\sum_{i=0}^{2g-2-(n+\tilde n)(2G-2)}\left(\lambda A\right)^i
\omega_{g,G}^{n,\tilde n,i},
\label{eestr}
\ee
where $\omega_{g,G}^{n,\tilde n,i}$ are calcullable coefficients independent
of $A$.  The expansion is reminiscent of partition functions of string theory,
with $1/N$ being the string coupling constant, $\lambda$ the string tension,
$g$ the worldsheet genus, and $n$ ($\tilde n$) the number of
orientation-preserving (-reversing) sheets of the map from the worldsheet to
the target.

The authors of [2,3] interpreted the coefficients of this expansion in terms
of rules for counting specific maps between surfaces.  Maps that are allowed
to contribute to the sum are branched covers of the target, with collapsed
handles and infinitesimal tubes that connect various sheets of the cover
(see [2,3] for details).  The string partition function can then be rewritten
as an integral over the moduli space $\CM$ of such coverings of $M$,
\be
\CZ=\int_\CM\der\nu\, W(\nu ),
\ee
with the location of the branch points, collapsed handles and connecting
tubes serving as coordinates on $\CM$.  The density $W(\nu )$ to be integrated
is given essentially by the symmetry factor of the cover (see [2,3]).

The structure of the $1/N$ expansion and the character of maps that contribute
to the rules of [2-4] give us some indications about what we can expect from
the QCD string theory.  There are no terms in (\ref{eestr}) with both $n$ and
$\tilde n$ equal to zero, which indicates that folds of the maps are
suppressed and the QCD string is sensitive to its ``extrinsic curvature'' in
the target.  This point of view is also supported by the fact that worldsheets
of genus zero do not contribute to $\CZ$, and by the form of the terms
exponential in the area.  The fact that both orientation-preserving and
orientation-reversing covers contribute indicates that the corresponding
string theory is non-chiral.

Notice also that the string theoretical expression (\ref{eestr}) for the QCD
partition function allows us to take the limit $\lambda\rightarrow 0$.
Although this limit is rather singular from the spacetime point of view, the
large-$N$ expansion serves as a nice regulator.  Indeed, when we take the
$\lambda\rightarrow 0$ limit for fixed $n$ and $\tilde n$ in (\ref{eestr}), we
obtain finite results.  From the point of view of string theory, this
corresponds to the zero tension limit, and the only structure carried by the
allowed maps are the so-called $\Omega$-points [3].  It seems very reasonable
to expect that in the zero tension limit, the string theory of 2D QCD is
topological.

\vskip.2truein
\centerline{\bf 3. Harmonic Topological Sigma Models}
\vskip.1truein

In the previous section we have seen that the coefficients of the large-$N$
expansion of the QCD partition function on an arbitrary spacetime surface
$M$ has the structure of a sum over classes of maps from the worldsheet to the
spacetime.  The coefficients of this sum are exponentials of the area of the
worldsheet that covers the spacetime times finite polynomials in the area.
One of the first observations made by the authors of [3] is that the
contribution to the leading term in this polynomial is essentially one from
each homotopy sector that is allowed to contribute by Kneser's formula.  As a
first step towards the worldsheet Lagrangian for QCD string theory in two
dimensions, let us first try to construct a string theory whose partition
function is essentially equal to one in each allowed homotopy sector of maps.

The field configurations of the theory are given solely by the maps from
a worldsheet $\Sigma$ (of genus $g$) to the spacetime $M$ (of genus $G$),
\be
\Phi:\Sigma\rightarrow M.
\ee
We will be using coordinates $X^\mu$ on the target, and $\sigma^a$ on the
worldsheet.  Before specializing to two target dimensions, we will work with
targets of arbitrary real dimension $D$.

The theory that we are about to construct is a topological sigma model, with
the topological gauge symmetry given by all possible deformations of $\Phi$.
We do not choose any other fixed geometrical structure either on the
worldsheet or in the target except the target metric $G_{\mu\nu}$.  In
particular, we do not assume a complex structure on $M$.  (In fact, the theory
will make perfect sense even for targets on which no complex structure
exists.)  The gauge-invariant Lagrangian is identically equal to zero, and we
use the standard BRST \index{BRST symmetry} technology to get a gauge-fixed
version of the theory.  First, we introduce the BRST charge that maps the
fields to their corresponding ghosts,
\be
[Q,X^\mu ]=\psi^\mu,\qquad\qquad\{Q,\psi^\mu\}=0.
\ee
Then we choose a gauge fixing condition.  Whereas the standard choice in the
traditional topological sigma model is the holomorphicity of $\Phi$, here we
will proceed in a different way.  Inspired by the results of the $1/N$
expanison of 2D QCD as reviewed in \S{2}, we are looking for a non-chiral
topological sigma model, and we need a gauge fixing condition that prefers
neither holomorphic nor anti-holomorphic maps.

To obtain a gauge fixing condition that satisfies these requirements, we pick
an auxiliary fixed metric $h_{ab}$ on the worldsheet, define the Laplacian on
the maps from the worldsheet to the target in the standard way:
\be
\Delta X^\mu\equiv h^{ab}\nabla_a\p_bX^\mu\equiv h^{ab}\left(\p_a\p_bX^\mu
+\christ\mu\sigma\rho\p_aX^\sigma\p_bX^\rho -\christ cab\p_cX^\mu\right),
\ee
and take the harmonicity condition
\be
\Delta X^\mu=0
\label{eeharm}
\ee
as the gauge fixing condition.%
\footnote{It is amusing to note here that the theory of harmonic maps between
surfaces (see [9] for details) produces one of the shortest and most elegant
analytic proofs of the (purely topological) Kneser formula that seems to be so
important for the 2D QCD string.}

To implement the gauge fixing, we introduce the antighost/auxiliary BRST
multiplet,
\be
\{Q,\chi^\mu\}=B^\mu,\qquad\qquad [Q,B^\mu]=0.
\ee
The tensorial properties of the multiplet are determined by the properties of
the gauge fixing condition (\ref{eeharm}), which in our case is a section of
$\Phi^{-1}(TM)$.  Note that both the ghosts and the antighosts are (fermionic)
sections of the same bundle over the worldsheet, an important fact that makes
the theory very different from the traditional topological sigma model.

The choice of the gauge fixed Lagrangian that ensures its general covariance
and yields the required gauge fixing condition is
\be
\CL=\ints\rth\{Q,G_{\mu\nu}\chi^\mu\left(-\frac{1}{2}B^\nu+\Delta X^\nu
+\frac{1}{2}\christ\nu\sigma\rho\chi^\sigma\psi^\rho\right)\}.
\ee
After performing the BRST commutator and integrating out the auxiliaries, we
end up with the following form of the gauge fixed Lagrangian:
\ba
\CL&=&\int_\Sigma\der^2\sigma\,\sqrt{h}\left(\frac{1}{2}G_{\mu\nu}\Delta X^\mu
\Delta X^\nu-G_{\mu\nu}\chi^\mu\Delta\psi^\nu+R_{\mu\sigma\rho\nu}h^{ab}\p_a
X^\sigma\p_bX^\rho\chi^\mu\psi^\nu\right.\label{eehtsm}\\
&&\qquad\qquad\qquad\qquad\qquad\qquad{}\left.-\frac{1}{2}
R_{\mu\sigma\rho\nu}\chi^\mu\psi^\sigma\chi^\rho\psi^\nu\right),\nonumber
\ea
which is invariant under the on-shell BRST transformation given by
\be
[Q,X^\mu ]=\psi^\mu ,\quad\{ Q,\psi^\mu\} =0,\quad\{Q,\chi^\mu\} =\Delta
X^\mu +\christ\mu\sigma\rho\chi^\sigma\psi^\rho.
\ee
In (\ref{eehtsm}) we have used the following definition of the covariant
Laplacian on the ghosts,
\ba
\Delta\psi^\mu=h^{ab}\nabla_a\nabla_b\psi^\mu&\equiv &h^{ab}
\left( \p_a\p_b\psi^\mu +2\frac{}{}\christ\mu\sigma\rho\p_aX^\sigma\p_b
\psi^\rho-\christ cab\p_c\psi^\mu\right)\\
&&\ {}+\frac{}{}\christ\mu\sigma\rho\Delta X^\sigma\psi^\rho +h^{ab}\p_a
X^\sigma\p_bX^\rho\psi^\nu\left(R^\mu{}_{\rho\sigma\nu}+\frac{}{}\p_\nu
\christ\mu\sigma\rho\right);\nonumber
\ea
in this notation, the BRST transformation of $\Delta X^\mu$ is equal to
$\Delta\psi^\mu$ plus additional terms,
\be
[Q,\Delta X^\mu ]=\Delta\psi^\mu -R^\mu{}_{\sigma\rho\lambda}h^{ab}\p_a
X^\sigma\p_bX^\rho\psi^\lambda -\christ\mu\sigma\rho\Delta X^\sigma\psi^\rho .
\label{eetrans}
\ee
This formula will be useful below.

The Lagrangian (\ref{eehtsm}) is an exact BRST commutator, and the path
integral of the theory is by standard arguments localized at the moduli space
of instantons, which in our case are harmonic maps from $\Sigma$ to $M$.  To
distinguish the theory from the traditional topological sigma model, I refer
to it as the ``harmonic topological sigma model.''

The Lagrangian of the harmonic topological sigma model resembles closely the
Lagrangian of topological mechanics.  First of all, since both the ghost field
$\psi^\mu$ and the antighost $\chi^\mu$ are sections of the same bundle,
their zero modes (in a fixed harmonic background $X^\mu_0$) satisfy identical
equations,
\be
\Delta\psi^\mu - R^\mu{}_{\sigma\rho\nu}h^{ab}\p_a X^\sigma_0\p_b X^\rho_0
\psi^\nu =0\label{eejac}
\ee
(and similarly for $\chi^\mu$), and the ghost number anomaly in the path
integral is identically zero.

The absence of quantum anomaly in the ghost number conservation has important
implications for the structure of correlation functions.  Topological sigma
models have typically two classes of observables.  The de~Rham cohomology
ring of the target produces one of them (with the degree of the cohomology
class being the ghost number of the observable).  If the first homotopy group
is non-trivial, there is another class of observables, corresponding
to the vacua of the winding sectors (cf.\ [10]).  Assuming that there are no
point-like observables of negative ghost numbers, every correlation function
that contains at least one observable of non-zero ghost number is zero by
the ghost number conservation in the quantum theory.

Given a fixed harmonic map $\Phi_0:\Sigma\rightarrow M$, we can analyze the
local structure of the moduli space of harmonic maps around $\Phi_0$ as
follows.  Consider a small deformation $\delta X^\mu$ of $X^\mu$.  The
deformed map is harmonic (to lowest order) if $\delta X^\mu$ satisfies the
linearized harmonicity condition,
\be
\Delta \delta X^\mu-\,R^\mu{}_{\sigma\rho\nu}h^{ab}\p_aX^\sigma_0\p_bX^\rho\,
\delta X^\nu=0.
\ee
This equation, also known in the theory of harmonic maps as the Jacobi
equation, is identical to Eqn.~(\ref{eejac}) for the zero modes of
$\psi^\mu$.  The (integrable) zero modes of $\psi^\mu$ are thus tangent to the
moduli space of harmonic maps, and the number of linear independent integrable
solutions to the Jacobi equation measures the dimension of the moduli space.

Except for the higher order form of the kinetic terms, the only apparent
difference between the Lagrangian of the harmonic sigma model and the
Lagrangian of topological mechanics is the existence of the
curvature-dependent two-fermi term in (\ref{eehtsm}).  One might na\"\i vely
expect that this term will join the curvature-dependent four-fermi term in
saturating the fermionic integral over the zero modes.  The BRST
transformation of $\Delta X^\mu$, Eqn.~(\ref{eetrans}), indicates that this is
not the case:  In the path integral, the curvature-dependent two-fermi term
gets absorbed into the kinetic term of the fermi fields, and leads to the
determinant cancelation in the (exact) Gaussian approximation.  Hence, the
four-fermi term remains the only one that can be used to saturate the
ghost/antighost zero modes in the path integral.

The structure of the partition function of the harmonic sigma model is thus
very similar to that of topological mechanics with the same target manifold:
In the homotopically trivial sector, harmonic maps are constant maps, and the
moduli space is isomorphic to the target.  The partition function is equal to
the Euler character of the target manifold.  Non-trivial homotopy sectors
provide interesting stringy corrections to this result, but what is always
computed is the Euler character of the moduli space of harmonic maps in the
corresponding homotopy class.

The fact that the path integral gives the Euler character of the moduli space
of harmonic maps can be proven rigorously, following Atiyah and Jeffrey [11].
The path integral of a given topological field theory computes a regularized
Euler number of a specific vector bundle $\CV$ over the infinite dimensional
manifold $\CA$ of all field configuations.  In our case, $\CA$ contains all
maps from $\Sigma$ to $M$, and $\CV$ is the bundle whose typical section is
the left hand side of the gauge fixing condition, i.e.\ $\CV$ is the tangent
bundle to $\CA$.  The regularized Euler number of this infinite dimensional
bundle is defined as the Euler number of a restriction of $\CV$ to the zero
locus of $\Delta X^\mu$, and coincides with the Euler character of the moduli
space of harmonic maps.

\vskip.1truein
\underline{Harmonic topological sigma models in two dimensions}
\vskip.1truein

Let us now consider, as an example, the harmonic topological sigma model with
a two dimensional target $M$ of genus $G$.  We know that the path integral of
the theory is localized at the moduli space of the harmonic maps from $\Sigma$
to $M$, and we would like to check explicitly that it computes the Euler
character of the moduli space.  In the case of maps between surfaces, harmonic
maps have been analyzed thoroughly in the mathematical literature (see [9] and
references therein), and many important results have been obtained.  Consider
for instance an arbitrary homotopy class of maps between surfaces with both
$g$ and $G$ greater than one, and with non-zero degree.  Then there is a
theorem [9] claiming that in such a homotopy class, there is exactly one
harmonic map!  The moduli space thus consists of only one point, there are no
zero modes of the fermi fields, and the path integral in this homotopy class
gives exactly one.  Analogously, in the homotopically trivial sector, the only
harmonic maps are the constant maps, and the moduli space coincides with $M$.
There are two zero modes for both $\psi^\mu$ and $\chi^\mu$, corresponding to
translations of the image of $\Phi$ in $M$.  These zero modes bring down the
curvature in the path integral, which is then equal to the Euler character of
the target.

The path integral is similarly simple in the general case with any two
dimensional target (see [5]).  The moduli space of harmonic maps between
surfaces with fixed metrics in a given homotopy class is always non-empty if
$G\geq 1$ [9]; the only case when there are no harmonic maps in some homotopy
classes corresponds to $G$ equal to zero, i.e.\ to the spherical target.

Up to this point we have studied the topological sigma model as a matter
system on the worldsheet.  To get a string theory instead, we have to take
care of worldsheet diffeomorphisms.  The minimal way of doing so is to take
the partition function of the harmonic topological sigma model, and sum over
all possible homotopy classes.  The sum is na\"\i vely infinite, and we have
to factorize by the group of global diffeomorphisms of the worldsheet.  The
resulting theory is finite, but it is still far from describing the QCD string
in two dimensions.

\vskip.2truein
\centerline{\bf 4. Topological Rigid Strings}
\vskip.1truein

In the standard way of gauging worldhseet diffeomorphisms in a topological
matter system, one introduces a topological gravity multiplet and couples it
to the matter system.  Here we will proceed in a different way, motivated
mainly by the observation that QCD string theory is probably very different
from traditional string theory.  In particular, the QCD string can be
sensitive to the extrinsic geometry, and the worldsheet quantization of
the string theory could use a spacetime cutoff instead of the worldsheet
cutoff, as e.g.\ in the rigid string theory.  The na\"\i ve dimension of
$X^\mu$ in such a theory is one instead of zero, and the quantum properties of
the string are governed by a fixed point that is substantially different from
the one usually employed in traditional critical string theory.
Unfortunately, such a string theory is very hard to quantize explicitly, and
little is known about its properties.

Here we will see how to avoid these shortcomings in a topological version of
such a string theory.  Indeed, we have noticed above that the bosonic part
of the Lagrangian for the harmonic topological sigma model is reminiscent
of the rigid string theory studied by Polyakov and others some time ago.
The basic discrepancy is that we have used a fixed fiducial metric on the
worldsheet (which does not violate the na\"\i ve background independence of
the theory by virtue of the topological symmetry), whereas the rigid string
theory uses the induced metric.  We will use this observation here to
gauge worldsheet diffeomorphisms as follows.  The BRST multiplet combines
the topological symmetry of the harmonic topological sigma model with the
worldsheet diffeomorphism symmetry,
\ba
[Q,X^\mu]&=&\psi^\mu+c^a\p_aX^\mu,\nonumber\\
\{Q,\psi^\mu\}&=&c^a\p_a\psi^\mu+\phi^a\p_aX^\mu,\\
\{ Q,c^a\}&=&c^b\p_bc^a-\phi^a,\nonumber\\
{[}Q,\phi^a]&=&c^b\p_b\phi^a-\phi^b\p_bc^a.\nonumber
\ea
Here $c^a$ is the diffeomorphism ghost, which can be omitted from the
(equivariant) theory, under the assumption that we deal exclusively with
diffeomorphism invariant objects on which the reduced BRST charge is still
nilpotent.  We have also introduced the bosonic ghost-for-ghost $\phi^a$ of
ghost number two, which takes care of the overcounting of worldsheet
diffeomorphisms in the BRST symmetry.

The gauge fixing condition in the bosonic sector remains formally the same
as in the harmonic topological sigma model,
\be
\rth\Delta X^\mu=0,
\label{eeminimal}
\ee
but now we have not introduced any fiducial metric on the worldsheet.
Instead, we identify the metric $h_{ab}$ with the induced metric,
\be
h_{ab}=\p_a X^\mu\p_b X^\nu\, G_{\mu\nu}.
\ee
Consequently, the Laplacian that enters the gauge fixing condition
(\ref{eeminimal}) is now given by
\be
\Delta X^\mu\equiv h^{ab}\left( \delta^\mu_\nu-h^{cd}G_{\nu\lambda}\p_cX^\mu
\p_dX^\lambda\right)\left(\p_a\p_bX^\nu+\christ\nu\sigma\rho\p_aX^\sigma\p_b
X^\rho\right).
\label{eeindlap}
\ee

One may wonder whether the na\"\i ve counting of gauge fixing conditions
agrees with the number expected in a diffeomorphism invariant theory.  While
in the harmonic topological sigma model in $D$ target dimensions the
harmonicity condition gives $D$ gauge fixing conditions (as it should), here
the requirement that $h_{ab}$ be the induced metric effectively reduces
the number of independent components of $\Delta X^\mu$ by two.  Indeed,
assuming that the induced metric is smooth and non-degenerate, $\Delta X^\mu$
coincides with the trace of the second fundamental form of the map
$\Phi:\Sigma\rightarrow M$, and it is easy to show that $\Delta X^\mu$ of
(\ref{eeindlap}) is normal to the image $\Phi(\Sigma)$ of $\Sigma$ in $M$,
\be
G_{\mu\nu}\,\Delta X^\mu\,\p_a X^\nu=0.
\label{eeortho}
\ee
In $D$ target dimensions, the number of independent components in
(\ref{eeminimal}) is thus $D-2$, which leaves exactly two unfixed coordinates
for worldsheet diffeomorphism symmetry.  In two dimensions, we seem to have no
condition at all. What makes the theory non-trivial, however, is the fact
that the requirements imposed on the induced metric that lead to
(\ref{eeortho}) are generically impossible to satisfy everywhere on the
worldsheet.  In particular, the gauge fixing condition forbids the existence
of folds on the map from $\Sigma$ to $M$.

The gauge fixing procedure can be easily completed at higher ghost numbers,
and we end up with a topological string Lagrangian of the following form:
\be
\CL=\ints\rth\left(\frac{}{}G_{\mu\nu}\Delta X^\mu\Delta X^\nu +{\rm ghost\
terms}\right).
\label{eetrslag}
\ee
Unlike in the harmonic topological sigma model, $h_{ab}$ in (\ref{eetrslag})
is now the induced metric.  The exact form of the ghost terms is really not so
important for our purposes here, and can be found elsewhere [5].  Indeed,
since the theory has been constructed as a topological string theory, we
do not need to know the exact form of the Lagrangian in order to compute the
partition function.  In topological theories, we know a priori what the path
integral is computing, and we can frequently do the calculation without
invoking the path integral.

In the case at hand, the Lagrangian (\ref{eetrslag}) defines a a topological
version of the rigid string theory.  The ghost-independent part of the
Lagrangian coincides with the rigid string theory Lagrangian of [6], and it
is natural to refer to the theory as the ``topological rigid string.''  Its
path integral is localized at the moduli space of solutions to the gauge
fixing solution $\sqrt{h}\Delta X^\mu=0$.  This means that the maps dominating
the path integral are harmonic in their own induced metric, i.e.\ the path
integral is localized at the moduli space of minimal-area maps, counted up to
worldsheet diffeomorphisms!

In the topological sigma model of \S{3}, we knew a priori that the path
integral computed the Euler number of the moduli space of harmonic maps, and
we confirmed this fact by a direct calculation for the simplest case of two
dimensional targets.  In the topological rigid string theory, similarly, we
know that the path integral gives an (equivariant) Euler number of the moduli
space of minimal maps.  The moduli spaces of minimal maps can be conveniently
parametrized if we adopt the spacetime point of view.  Indeed, the
minimal-area condition forbids folds of the maps, and only allows branch
points (and possibly collapsed handles).  The location of branch points
in the spacetime then serves as a natural set of coordinates on the moduli
space of minimal-area maps.  The branch points are indistinguishable, as a
result of the worlsheet diffeomorphism symmetry.  Consequently, the moduli
spaces that dominate the path integral of the topological rigid string are
quite similar to the spaces of maps that contribute to the $1/N$ expansion of
2D QCD.  More details about this correspondence can be found in [5].

What we have considered thus far is a topological string theory with zero
tension.  Its partition function is naturally independent of the area of the
target.  To obtain a string theory with non-zero tension, we would like to add
to the Lagrangian a term that behaves as
\be
\delta\CL=\lambda\ints\rth.
\label{eearealag}
\ee
When evaluated at the maps that dominate the path integral of the topological
rigid string theory, i.e.\ at the minimal-area maps, this term gives
\be
\delta\CL (\Phi)=\lambda(n+\tilde n)A,
\ee
where $n,\tilde n$ are determined by the homotopy class of the minimal map
$\Phi$ and represent the number of orientation-preserving and
orientation-reversing sheets of the worldsheet over the spacetime, and $A$ is
the area of the target.  If we add (\ref{eearealag}) to the Lagrangian of the
topological rigid string and then evaluate the deformed Lagrangian at the
moduli space, we recover exactly the same exponential dependence on the area
as observed in 2D QCD.

Of course, this cannot be the whole story.  The na\"\i ve area term
(\ref{eearealag}) does not respect the topological BRST symmetry of the
rigid string theory and cannot be added to the Lagrangian without spoiling its
BRST invariance.  We can, however, make the induced area BRST-invariant
by adding ghost-dependent terms to (\ref{eearealag}):
\be
\delta\CL'=\lambda\ints\left(\rth+{\rm ghost\ terms}\right),\qquad {[}Q,
\delta\CL']=0.
\ee
This term can be added to the topological rigid string Lagrangian.  In the
partition function, the ghost corrections to the area term do not change the
form of the terms that are exponential in the area.  They do affect, however,
the integral over the moduli space of minimal-area maps.  Indeed, in the path
integral of the deformed theory, the ghost corrections to (\ref{eearealag})
will enter the integral over the fermionic zero modes, thus changing the
integration over the bosonic zero modes.  In the partition function, this will
produce a polynomial dependence on the target area in each homotopy sector of
maps from the worldsheet to the target.

This procedure is reminiscent of the mechanism that connects the ``physical''
Yang-Mills theory in the target to a topological version thereof (cf.\ \S{2}).
It would be very interesting to see whether a similar mechanism can be
involved in the worldsheet theory of the 2D QCD string as well.

\vskip.2truein
\centerline{\bf 5. Higher Dimensions}
\vskip.1truein

In previous paragraphs, we have presented some evidence in favor of the rigid
string picture of QCD, in the exactly solvable case of the two dimensional,
pure-glue theory on compact surfaces.  We have observed some striking
similarities between a topological rigid string theory and the results of
the large-$N$ expansion in QCD as obtained in [2-4].  Here I comment on
possible extensions of the results to higher dimensions.

Although both the harmonic topological sigma model and the topological rigid
string theory can be in principle constructed in arbitrary real dimension, the
four-dimensional case is unique.  Let us first rewrite the rigid string
Lagrangian
\be
\CL=\ints\rth G_{\mu\nu}\Delta X^\mu\Delta X^\nu
\ee
in an equivalent form,
\be
\CL=\ints\rth h^{ab}\nabla_a t^{\mu\nu}\nabla_b t^{\sigma\rho}G_{\mu\sigma}
G_{\nu\rho},
\label{eerigalt}
\ee
with $t^{\mu\nu}$ defined by
\be
t^{\mu\nu}=\frac{\epsilon^{ab}}{\rth}\p_aX^\mu\p_bX^\nu.
\ee
In four dimensions, the rigid string theory described by (\ref{eerigalt}) is
known to have instantons, with the instanton number being the
self-intersection number $s(\Phi )$ of the worldsheet in the target.
It is easy to show that
\be
s(\Phi )=\ints\epsilon^{ab}\nabla_a t^{\mu\nu}\nabla_b t^{\sigma\rho}
\epsilon_{\mu\nu\sigma\rho}.
\label{eeinst}
\ee
In the string picture of 4D QCD, $s(\Phi )$ is believed to correspond to
the $\theta$ angle of the spacetime Yang-Mills theory [12].  It is possible to
write down a Lagrangian for a topological rigid string theory in four
dimensions, such that its path integral is localized at the moduli space of
instantons with the instanton number given by (\ref{eeinst}).  One can
naturally ask whether such a topological rigid string theory would correspond
to topological Yang-Mills theory in four dimensions [13], i.e.\ to the
$\su{N}$ Donaldson theory in $1/N$ expansion.  This question should be much
easier to answer that its full-fledged physical counterpart.

\vskip.2truein
\centerline{\bf Acknowledgement}
\vskip.1truein

It is a pleasure to thank the organizers for creating such a stimulating
environment in Carg\`ese and for the opportunity to present these results.
During the course of the work I have benefitted from discussions with and
comments from P. Freund, D. Gross, D. Kutasov, E. Martinec and E. Witten.
The research has been supported in part by NSF Grant PHY90-00386 and DOE Grant
DEFG02-90ER40560.

\vskip.2truein
\centerline{\bf References}

\begin{enumerate}
\item{D.J. Gross, ``Some New/Old Approaches to QCD,''
Berkeley/Princeton preprint LBL-33232=PUPT-1355 (November 1992)\endli
J. Polchinski, ``Strings and QCD?,'' Austin preprint UTTG-16-92 (June 1992)}
\item{D.J. Gross, ``Two Dimensional QCD as a String Theory,''
\npb{400}{93}{161}}
\item{D.J. Gross and W. Taylor, ``Two Dimensional QCD is a String Theory,''
\npb{400}{93}{181}; ``Twists and Wilson Loops in the String Theory of Two
Dimensional QCD,'' \npb{403}{93}{395}}
\item{V.A. Kazakov and I.K. Kostov, ``Non-Linear Strings in Two Dimensional
U($\infty$) Gauge Theory,'' \npb{176}{80}{199}\endli
V.A. Kazakov, ``Wilson Loop Average for an Arbitrary Contour in
Two-Dimensional U($N$) Gauge Theory,'' \npb{179}{81}{283}\endli
for recent reviews, see:\endli
V. Kazakov, ``A String Project in Multicolour QCD,'' LPTENS preprint (1993),
\endli
I.K. Kostov, ``$U(N)$ Gauge Theory and Lattice Strings,'' Saclay preprint
T93/079 (August 1993),\endli
and this volume}
\item{P. Ho\v rava, ``Topological Rigid String Theory and Two Dimensional
QCD,'' Chicago preprint, to appear (November 1993)}
\item{A.M. Polyakov, ``Fine Structure of Strings,'' \npb{268}{86}{406};
``{\sl Gauge Fields and Strings},'' \S{10.4} (Harwood Publishers, 1987);\endli
``A Few Projects in String Theory,'' Princeton preprint PUPT-1394 (April 1993)}
\item{A.A. Migdal, ``Recursion Equations in Gauge Field Theories,''
{\it Zh.\ Eksp.\ Teor.\ Fiz.} {\bf 69} (1975) 810 [{\it Sov.\ Phys.\ JETP}$\,$
{\bf 42} (1975) 413]\endli
B.Ye.\ Rusakov, ``Loop Averages and Partition Functions in
$\u{N}$ Gauge Theory on Two-Dimensional Manifolds,'' \mpla{5}{90}{693};
``Large-N Quantum Gauge Theories in Two Dimensions,'' Tel-Aviv preprint
TAUP-2012-92 (December 1992)}
\item{E. Witten, ``On Quantum Gauge Theories in Two Dimensions,''
\cmp{141}{91}{153}; ``Two Dimensional Gauge Theories Revisited,'' {\it J.
Geom.\ Phys.} {\bf 9} (1992) 303}
\item{J. Eells and L. Lemaire, ``A Report on Harmonic Maps,'' {\it Bull.\
London Math.\ Soc.} {\bf 10} (1980) 1; ``Another Report on Harmonic Maps,''
{\it Bull.\ London Math.\ Soc.} {\bf 20} (1988) 385; ``On the Construction of
Harmonic and Holomorphic Maps between Surfaces,'' {\it Math.\ Ann.}
{\bf 252} (1980) 27}
\item{P. Ho\v rava, ``Two Dimensional String Theory and the Topological
Torus,'' \npb{386}{92}{383}; ``Spacetime Diffeomorphisms and Topological
$\winfty$ Symmetry in Two Dimensional Topological String Theory,'' Chicago
preprint EFI-92-70 (January 1993) hep-th/9302020, to appear in {\it Nucl.\
Phys.} {\bf B} (1993)}
\item{M.F. Atiyah and L. Jeffrey, ``Topological Lagrangians and Cohomology,''
{\it J. Geom.\ Phys.} {\bf 7} (1990) 119\endli
for a review directed to physicists, see:\endli
M. Blau, ``The Mathai-Quillen Formalism and Topological Field Theory,''
Amsterdam preprint NIKHEF-H/92-07 (March 1992)}
\item{P.O. Mazur and V.P. Nair, ``Strings in QCD and $\theta$ Vacua,''
\npb{284}{86}{146}}
\item{E. Witten, ``Topological Quantum Field Theory,'' \cmp{117}{88}{353}}
\end{enumerate}
\end{document}